\RequirePackage{fix-cm}
\documentclass[smallextended]{svjour3}       
\smartqed  
\usepackage{graphicx}
\usepackage{amsmath,amssymb}
\usepackage[english]{babel}
\usepackage{layout}
\usepackage{enumerate}
\usepackage{cancel}
\usepackage{bm}

\usepackage{hyperref}\hypersetup{
    colorlinks,%
    citecolor=blue,%
    filecolor=blue,%
    linkcolor=blue,%
    urlcolor=blue
}

\begin{document}

\title{Erratum to: Geodesic Deviation Equation in $\bm{f(R)}$ Gravity}

\author{Alejandro Guarnizo \and Leonardo Casta\~neda \and Juan M. Tejeiro}

\institute{A. Guarnizo \at Institut F\"{u}r Theoretische Physik, Ruprecht-Karls-Universit\"{a}t Heidelberg,\\
           Philosophenweg 16, 69120 Heidelberg, Germany\\
            \email{a.guarnizo@thphys.uni-heidelberg.de}
	    \and 
	    L. Casta\~neda \at
	    Observatorio Astron\'omico Nacional,\\
	    Universidad Nacional de Colombia, Bogot\'a, Colombia\\
	    \email{lcastanedac@unal.edu.co}
	    \and 
	    J. M. Tejeiro \at 
	    Observatorio Astron\'omico Nacional,\\
	    Universidad Nacional de Colombia, Bogot\'a, Colombia\\
	    \email{jmtejeiros@unal.edu.co}
}

\date{Received: date / Accepted: date}

\maketitle

\begin{abstract}
In the context of metric $f(R)$ gravity, the Geodesic Deviation Equation (GDE) was first studied in \cite{Guarnizo:2010ng}, 
giving a general expression and studying a particular case, the FLRW universe.  In
the paper \cite{delaCruz-Dombriz:2013gfa} a similar
analysis was made. However, there is a discrepancy in the expressions for the null
vector field case due to an algebraic error in our original paper. Here, we make explicit
the contribution of the different operators in the GDE, and we correct our previous
results.
\end{abstract}

\section{Erratum to: Gen Relativ Gravit (2011) 43:2713-2728 \\ \href{http://link.springer.com/article/10.1007/s10714-011-1194-6}{DOI 10.1007/s10714-011-1194-6}}
The general expression for the GDE in $f(R)$ gravity is given by \cite[eq. (10)]{Guarnizo:2010ng}. The r.h.s of the GDE could be written as

\begin{multline}
R_{\beta\gamma\delta}^{\alpha}V^{\beta}\eta^{\gamma}V^{\delta} = C_{\beta\gamma\delta}^{\alpha}V^{\beta}\eta^{\gamma}V^{\delta}
+ \frac{1}{2f'}\Biggl[\kappa(T_{\delta\beta}\delta_{\gamma}^{\alpha}-T_{\gamma\beta}\delta_{\delta}^{\alpha}
+ T_{\gamma}^{\, \, \alpha}g_{\beta\delta}-T_{\delta}^{\, \, \alpha}g_{\beta\gamma})\\
+ f\bigl(\delta_{\gamma}^{\alpha}g_{\delta\beta}
- \delta_{\delta}^{\alpha}g_{\gamma\beta}\bigr)
+  \bigl(\delta_{\gamma}^{\alpha}\mathcal{D}_{\delta\beta}
 - \delta_{\delta}^{\alpha}\mathcal{D}_{\gamma\beta} + g_{\beta\delta}\mathcal{D}_{\gamma}^{\, \, \alpha}- g_{\beta\gamma}\mathcal{D}_{\delta}^{\, \, \alpha}\bigr)f'\Biggr]V^{\beta}\eta^{\gamma}V^{\delta}\\
 - \frac{1}{6f'}\biggl(\kappa T + 2f  - 3\square f'\biggr)\bigl(\delta_{\gamma}^{\alpha}g_{\delta\beta} - \delta_{\delta}^{\alpha}g_{\gamma\beta}\bigr)V^{\beta}\eta^{\gamma}V^{\delta},
\end{multline}
being $R_{\beta\gamma\delta}^{\alpha}$ the Riemann curvature tensor, $\eta^{\alpha}$ the deviation vector between geodesics of tangent vector field $V^{\alpha}$,
$\mathcal{D}_{\alpha\beta}\equiv g_{\alpha\beta}\square - \nabla_{\alpha}\nabla_{\beta}$, $\square  = \nabla_{\sigma}\nabla^{\sigma}$, $f' = f'(R) = df(R)/dR$, $\kappa = 8\pi G$, $T_{\alpha\beta}$ the energy-momentum tensor
and $T$ its trace. The contribution of the operators $\mathcal{D}_{\alpha\beta}$ could be further simplified as

\begin{multline}
\bigl(\delta_{\gamma}^{\alpha}\mathcal{D}_{\delta\beta}
 - \delta_{\delta}^{\alpha}\mathcal{D}_{\gamma\beta} + g_{\beta\delta}\mathcal{D}_{\gamma}^{\, \, \alpha}- g_{\beta\gamma}\mathcal{D}_{\delta}^{\, \, \alpha}\bigr)f'V^{\beta}\eta^{\gamma}V^{\delta}\\
 = (\nabla_{\delta}\nabla_{\beta}f')V^{\beta}\eta^{\alpha}V^{\delta}-2\epsilon(\square f')\eta^{\alpha}-(\nabla_{\gamma}\nabla_{\beta}f')V^{\beta}\eta^{\gamma}V^{\alpha}
 +\epsilon(\nabla_{\gamma}\nabla^{\alpha}f')\eta^{\gamma},
\end{multline}
with $\epsilon = V^{\alpha}V_{\alpha}$, and using $\eta_{\alpha}V^{\alpha}=0$. The r.h.s of the GDE reduces to 

\begin{multline}
R_{\beta\gamma\delta}^{\alpha}V^{\beta}\eta^{\gamma}V^{\delta} = C_{\beta\gamma\delta}^{\alpha}V^{\beta}\eta^{\gamma}V^{\delta}
+ \frac{1}{2f'}\Biggl[\kappa\bigl(T_{\delta\beta}V^{\beta}\eta^{\alpha}V^{\delta}-T_{\gamma\beta}V^{\beta}\eta^{\gamma}V^{\alpha}
+ \epsilon T_{\gamma}^{\, \, \alpha}\eta^{\gamma}\bigr)\\
- \epsilon\biggl(\frac{\kappa T}{3} - \frac{f}{3} + \square f'\biggr) \eta^{\alpha}
+  (\nabla_{\delta}\nabla_{\beta}f')V^{\beta}\eta^{\alpha}V^{\delta}-(\nabla_{\gamma}\nabla_{\beta}f')V^{\beta}\eta^{\gamma}V^{\alpha}\\
 +\epsilon(\nabla_{\gamma}\nabla^{\alpha}f')\eta^{\gamma} \Biggr].
\end{multline}
which is the general expression for any metric, and any energy-momentum content.
\subsection{Geodesic Deviation Equation for the FLRW universe}
In this particular case we have
\begin{equation}
T_{\alpha\beta} = (\rho + p)u_{\alpha}u_{\beta} +p g_{\alpha\beta}, \qquad T = 3p-\rho,
\end{equation}
the r.h.s. of the GDE becomes
\begin{multline}
R_{\beta\gamma\delta}^{\alpha}V^{\beta}\eta^{\gamma}V^{\delta} = \frac{1}{2f'}\Biggl[\biggl(\kappa(\rho +p)E^2 +
\frac{\epsilon}{3}\bigl(\kappa(\rho+3p) + f-3\square f'\bigr)\biggr) \eta^{\alpha}\\
+  (\nabla_{\delta}\nabla_{\beta}f')V^{\beta}\eta^{\alpha}V^{\delta}-(\nabla_{\gamma}\nabla_{\beta}f')V^{\beta}\eta^{\gamma}V^{\alpha}
 +\epsilon(\nabla_{\gamma}\nabla^{\alpha}f')\eta^{\gamma} \Biggr],
\end{multline}
with $E=-\eta_{\alpha}V^{\alpha}$, and $\eta_{\alpha}u^{\alpha}=0$. For the FLRW case the covariant derivatives are
\begin{align}
\nabla_0 \nabla_0 f' & = f''\ddot{R}+f'''\dot{R}^2, \\ \nonumber
\nabla_i \nabla_j f' & = -Hg_{ij}f''\dot{R}, \\ \nonumber
\square f' & = -f''(\ddot{R}+3H\dot{R})-f'''\dot{R}^2,
\end{align}
since in this case the four-velocity is $u^{\alpha}=(1,0,0,0)$ from the orthogonality conditions we get
$E=-V_{\alpha}u^{\alpha}=-V_0$, $\eta_{\alpha} u^{\alpha}=\eta_0u^0=0$ (thus, the deviation vector just have non-vanishing spatial
components $\eta^0 =0$), and $\eta_{\alpha}V^{\alpha}=\eta_{i}V^{i}$. Using these results and expanding explicitly

\begin{align}
(\nabla_{\delta}\nabla_{\beta}f')V^{\beta}V^{\delta} &=  (\nabla_{0}\nabla_{0}f')V^{0}V^{0} + (\nabla_{i}\nabla_{j}f')V^{i}V^{j},\notag\\
  &=  (f''\ddot{R}+f'''\dot{R}^2)E^2 - Hf''\dot{R}g_{ij}V^{i}V^{j},\notag\\
  &=  (f''\ddot{R}+f'''\dot{R}^2)E^2 - Hf''\dot{R}V_{j}V^{j},\notag\\
  &=  (f''\ddot{R}+f'''\dot{R}^2)E^2 - Hf''\dot{R}(\epsilon-V_0V^0),\notag\\
   &=  (f''\ddot{R}+f'''\dot{R}^2)E^2 - Hf''\dot{R}(\epsilon+E^2),
\end{align}
\begin{align}
(\nabla_{\gamma}\nabla_{\beta}f')V^{\beta}\eta^{\gamma}V^{\alpha} &=  (\nabla_{0}\nabla_{0}f')V^{0}\cancelto{0}{\eta^{0}}V^{\alpha}
+ (\nabla_{i}\nabla_{j}f')V^{i}\eta^{j}V^{\alpha},\notag\\
&=  -Hf''\dot{R}g_{ij}V^{i}\eta^{j}V^{\alpha},\notag\\
&=  -Hf''\dot{R}\cancelto{0}{V_{j}\eta^{j}}V^{\alpha}, \notag \\
&=0
\end{align}
\begin{align}
(\nabla_{\gamma}\nabla^{\alpha}f')\eta^{\gamma} &=  (\nabla_{\gamma}g^{\alpha\sigma}\nabla_{\sigma}f')\eta^{\gamma},\notag\\
&=  g^{\alpha\sigma}(\nabla_{\gamma}\nabla_{\sigma}f')\eta^{\gamma},\notag\\
&=  g^{\alpha 0}(\nabla_{0}\nabla_{0}f')\cancelto{0}{\eta^{0}}+g^{\alpha j}(\nabla_{i}\nabla_{j}f')\eta^{i},\notag\\
&=  -Hf''\dot{R}g^{\alpha j}g_{ij}\eta^{i},\notag\\
&=  -Hf''\dot{R}\delta^{\alpha}_j\eta^{j},\notag\\
&= -Hf''\dot{R}\eta^{\alpha}.
\end{align}
Then we see that in comparison with our result \cite[eq. (32)]{Guarnizo:2010ng}, there is an additional contribution to $E^2$ 
in the operators, which is missing. The method used in \cite{delaCruz-Dombriz:2013gfa} relies in the $1+3$ 
decomposition, and for the FLRW it is equivalent to the $3+1$ where the four velocity field for threading the 
spacetime is the four-velocity of the fundamental observers. The r.h.s. of the GDE reduces to
\begin{multline}\label{eq:gde}
R_{\beta\gamma\delta}^{\alpha}V^{\beta}\eta^{\gamma}V^{\delta} = \frac{1}{2f'}\Biggl[\biggl(\kappa(\rho +p)+f''(\ddot{R}-H\dot{R})+f'''\dot{R}^2\biggr)E^2\\
+ \biggl(\frac{\kappa\rho}{3}+\kappa p + \frac{f}{3}+f''(\ddot{R}+H\dot{R})+f'''\dot{R}^2\biggr)\epsilon\Biggr] \eta^{\alpha}.
\end{multline}
Defining the following quantities
\begin{equation}
\rho_{\text{eff}} = \frac{1}{f'}\left[\kappa \rho + \frac{Rf'-f}{2}-3Hf''\dot{R}\right], 
\end{equation}
\begin{equation}
p_{\text{eff}} = \frac{1}{f'}\left[\kappa p + \frac{f-Rf'}{2}+f''(\ddot{R} + 2H\dot{R}) + f'''\dot{R}^2\right],
\end{equation}
equation (\ref{eq:gde}) could be written in a more compact form \cite{delaCruz-Dombriz:2013gfa}
\begin{equation}\label{eq:Riemeff}
R_{\beta\gamma\delta}^{\alpha}V^{\beta}\eta^{\gamma}V^{\delta} =  \frac{1}{2}\left[(\rho_{\text{eff}}+p_{\text{eff}})E^2   + \frac{1}{3}\left(\rho_{\text{eff}} + 3p_{\text{eff}} + R \right)\epsilon\right]\eta^{\alpha},
\end{equation}
and finally we can write the GDE as
\begin{equation}\label{GDeFR}
\frac{D^2 \eta^{\alpha}}{D \nu^2} = - \frac{1}{2}\left[(\rho_{\text{eff}}+p_{\text{eff}})E^2   + \frac{1}{3}\left(\rho_{\text{eff}} +3p_{\text{eff}} + R \right)\epsilon\right]\eta^{\alpha},
\end{equation}
with $\frac{D}{D\nu}$ the covariant derivative along the curve. 
\subsubsection{GDE for Fundamental Observers}
In this case $V^{\alpha}$ is the four-velocity of the fluid $u^{\alpha}$. The affine parameter $\nu$ matches with the proper time of the fundamental observer
$\nu = t$. Because we have temporal geodesics then $\epsilon=-1$ and also the vector field is normalized $E = 1$, thus from (\ref{eq:gde})

\begin{equation}\label{GDEfR1}
R_{\beta\gamma\delta}^{\alpha}u^{\beta}\eta^{\gamma}u^{\delta} = \frac{1}{f'}\biggl[\frac{\kappa\rho}{3}  -\frac{f}{6} -  H  f'' \dot{R}\biggr]\eta^{\alpha},
\end{equation}
if the deviation vector is $\eta_{\alpha} = \ell e_{\alpha}$, using that $e_{\alpha}$ is parallel propagated along $t$, $\frac{D e^{\alpha}}{D t} = 0$, and
$\frac{D^2 \eta^{\alpha}}{D t^2} = \frac{d^2\ell}{dt^2} e^{\alpha}$, putting these results in the GDE (\ref{GDeFR}) with (\ref{GDEfR1}) gives

\begin{equation}
\frac{d^2\ell}{dt^2} = - \frac{1}{f'}\biggl[\frac{\kappa\rho}{3}  -\frac{f}{6} -  H  f'' \dot{R}\biggr]\, \ell .
\end{equation}
In particular with $\ell = a(t)$: 
\begin{equation}\label{Raycha}
\frac{\ddot{a}}{a} = \frac{1}{f'}\biggl[\frac{f}{6}  + H f'' \dot{R} -\frac{\kappa\rho}{3} \biggr],
\end{equation}
which is the Raychaudhuri equation for $f(R)$ gravity. The standard form of the modified Friedmann equations in $f(R)$ gravity, could be obtained from this equation.
\subsubsection{GDE for Null Vector Fields}
Now, we consider the GDE for past directed null vector fields. 
In this case $V^{\alpha}=k^{\alpha}$, $k_{\alpha}k^{\alpha}=0$, then Eq. (\ref{eq:Riemeff}) reduces to\\
\begin{equation}\label{RicciFoc}
R_{\beta\gamma\delta}^{\alpha}k^{\beta}\eta^{\gamma}k^{\delta} = \frac{1}{2}(\rho_{\text{eff}}+p_{\text{eff}})E^2\,\eta^{\alpha},
\end{equation}\\
that could be interpreted as the \textit{Ricci focusing} in $f(R)$ gravity. Writing $\eta^{\alpha}= \eta e^{\alpha}$,  $e_{\alpha}e^{\alpha}=1$, $e_{\alpha}u^{\alpha}=e_{\alpha}k^{\alpha}=0$ and choosing an aligned base parallel propagated
$\frac{D e^{\alpha}}{D \nu}=k^{\beta}\nabla_{\beta}e^{\alpha}=0$, the GDE (\ref{GDeFR}) is\\
\begin{equation}\label{GDE3}
\frac{d^2\eta}{d\nu^2} = - \frac{1}{2}(\rho_{\text{eff}}+p_{\text{eff}})E^2\, \eta.
\end{equation}\\
Using the transformation between the affine parameter $\nu$ and the redshift $z$, $\frac{d}{d\nu} \longrightarrow \frac{d}{dz} $ 
\cite[eq. (54)]{Guarnizo:2010ng}, the equation (\ref{GDE3}) could be written as \cite{delaCruz-Dombriz:2013gfa}\\
\begin{equation}
\frac{d^2\eta}{dz^2} + \frac{(7+3w_{\text{eff}})}{2(1+z)}\, \frac{d\eta}{dz} + \frac{3(1+w_{\text{eff}})}{2(1+z)^2}\, \eta = 0,
\end{equation}\\
with $w_{\text{eff}}=\frac{p_{\text{eff}}}{\rho_{\text{eff}}}$. 
This equation gives for FLRW the angular diametral distance $D_A$\footnote{Since $\eta \propto D_A$.}:\\
\begin{equation}\label{Angdist}
\frac{d^2 \, D_A}{dz^2} + \frac{(7+3w_{\text{eff}})}{2(1+z)}\, \frac{d \, D_A}{dz} + \frac{3(1+w_{\text{eff}})}{2(1+z)^2}\, D_A = 0.
\end{equation}\\
This equation is exactly the same result shown by \cite{delaCruz-Dombriz:2013gfa}. 


\section{An \textit{Alternative} Derivation}
The results for the angular diametral distance could be also obtained from the focusing equation (see \cite{Guarnizo:2010ng} and references therein). The method give us the direct result in concordance with~(\ref{Angdist}) confirming the correct result:
\begin{equation}
\frac{d^2D_A}{d\nu^2}= -\left(|\sigma|^2+\frac{1}{2}R_{\alpha\beta}k^{\alpha}k^{\beta}\right)D_A,
\end{equation}
being $\sigma$ the shear and $k^{\alpha}$ a null vector. From the field equations in $f(R)$ gravity we can write
\begin{equation}
R_{\alpha\beta}  = \frac{1}{f'}\left[\kappa T_{\alpha\beta}+\frac{f}{2}g_{\alpha\beta} +\nabla_{\alpha}\nabla_{\beta}f'-g_{\alpha\beta}\square f' \right],
\end{equation}
and for the specific case of the FLRW universe ($\sigma = 0$), the previous expression gives
\begin{equation}
R_{\alpha\beta}k^{\alpha}k^{\beta}  = \frac{1}{f'}\left[\kappa(\rho+p) +f''(\ddot{R}-H\dot{R})+f'''\dot{R}^2 \right]E^2,
\end{equation}
and then 
\begin{equation}
\frac{d^2D_A}{d\nu^2}= -\frac{1}{2f'}\left[\kappa(\rho+p) +f''(\ddot{R}-H\dot{R})+f'''\dot{R}^2 \right]E^2D_A,
\end{equation}
which has the same form as equation (\ref{GDE3}) (again with $\eta \propto D_A$). 

\section{Conclusions and Discussion}
In this short remark we clarify the difference in the results, between \cite{Guarnizo:2010ng} and \cite{delaCruz-Dombriz:2013gfa}, 
for the GDE in the null vector field case due to an algebraic error in our original paper. 
As we expect, the expressions are consistent with the Null Energy Condition (NEC) in $f(R)$ gravity \cite{Alcaniz:2007}. \\\\
As a final comment, we should notice that the contributions coming from the first term of r.h.s.
of (\ref{eq:gde}) makes the Dyer-Roeder equation \cite[eq. (80)]{Guarnizo:2010ng} more involving than our original 
result, because the smoothness parameter $\alpha$ (or any function describing the clumpiness of the universe) now depends on the $f(R)$ function and its contribution to the density perturbation. Equations for the density perturbations in $f(R)$ gravity are coupled with perturbations in the Ricci scalar and it depends on the $f(R)$ function (see~\cite[eq.(7)]{Zhao:2011}) making the effect of the perturbations in the geometrical part of the Ricci-focusing  a relevalant issue that should be investigated from the analysis of the full set of the optical Sachs equations.


\begin{thebibliography}{}
%
%


\bibitem{Guarnizo:2010ng} Guarnizo, A., Castaneda, L., Tejeiro, J. M.: Geodesic deviation equation in $f(R)$ gravity.
Gen. Rel. Grav. \textbf{43}, 2713-2728 (2011). [arXiv:1010.5279]
\bibitem{delaCruz-Dombriz:2013gfa}  de la Cruz-Dombriz, A.,  Dunsby, P. K. S.,  Busti, V. C.,  Kandhai, S.: On tidal forces in $f(R)$ 
theories of gravity. Phys.Rev. D \textbf{89}, 064029 (2014). [arXiv:1312.2022].
\bibitem{Alcaniz:2007} Santos, J., Alcaniz, J. S., Rebou\c{c}as, M. J., Carvalho,  F. C.: Energy conditions in $f(R)$ gravity.
Phys. Rev. D \textbf{76}, 083513 (2007). [arXiv:0708.0411]
\bibitem{Zhao:2011} Zhao, G.B., Koyama, K.: $N-$Body simulations for $f(R)$ gravity using a self-adaptive particle-mesh code   .
Phys. Rev. D \textbf{83}, 044007 (2011). [arXiv:1011.1257]
\end{thebibliography}
\end{document}